\documentclass[prb,twocolumn]{revtex4-1}
\usepackage{graphicx}
\usepackage{xcolor}

\newcommand{\bk}{\mathbf{k}}
\newcommand{\bp}{\mathbf{P}}

\begin{document}
\title{Quantum Anomalous Hall Effect in a Perovskite and Inverse-Perovskite
Sandwich Structure}
\author{Long-Hua Wu$^{1,2}$} 
\email{Wu.Longhua@nims.go.jp}
\author{Qi-Feng Liang$^{1,3}$, and Xiao Hu$^{1,2}$}
\email{Hu.Xiao@nims.go.jp}
\affiliation{$^1$International Center for Materials Nanoarchitectonics (WPI-MANA),\\
National Institute for Materials Science, Tsukuba 305-0044, Japan \\
$^2$Graduate School of Pure and Applied Sciences, University of
Tsukuba, Tsukuba 305-8571, Japan\\
$^3$Department of Physics, Shaoxing University, Shaoxing 312000, China}

\begin{abstract}
Based on first-principles calculations, we propose a sandwich structure
composed of a G-type anti-ferromagnetic (AFM) Mott insulator LaCrO$_3$ grown
along the [001] direction with one atomic layer replaced by an inverse-perovskite
material Sr$_3$PbO. We show that the system is in a topologically nontrivial phase
characterized by simultaneous nonzero charge and spin Chern numbers, which
can support a spin-polarized and dissipationless edge current in a finite system.
Since these two materials are stable in bulk and match each other with only
small lattice distortions, the composite material is expected easy to synthesize.
\end{abstract}
\maketitle

\section{Introduction}
Discovery of the quantum Hall effect (QHE) by von Klitzing has opened a new era in
condensed matter physics \cite{Klitzing1980}. It is revealed that the quantization
of Hall conductance is a manifestation of topologically nontrivial Bloch
wavefunctions~\cite{Thouless1982}. Topological matters have various promising
applications in many fields, such as fault-tolerant topological quantum
computations \cite{Nayak2008,Stanescu2013,Beenakker2013,Wu2014,Kawakami2015},
spintronics \cite{Oleg2010,PesinNM} and photonics \cite{Haldane2008,Khanikaev2013,Wu2015}.

The quantum spin Hall effect (QSHE) was first predicted theoretically in graphene
\cite{Kane2005a,Kane2005b,RMPKane} and later studied in a two-dimensional (2D)
HgTe quantum well both theoretically \cite{Bernevig,RMPZhang} and
experimentally \cite{ExpQSHE}. A 3D topological insulator Bi$_{1-x}$Sb$_x$
and its family members were also reported
\cite{Hsieh2008,HJZhang2009,Ando2013}. Breaking time-reversal symmetry can
drive a topological insulator into the quantum anomalous Hall effect (QAHE)
\cite{Haldane1988,Weng2015}. There are two categories of the QAHE classified
by the spin Chern number \cite{Prodan2009,DNSheng2011}. One subclass of the
QAHE is characterized by a vanishing spin Chern number. The Cr-doped
Bi$_2$Se$_3$ thin film \cite{Weng2015,Yu2010,ExpQAHE} belongs to this class,
where the topological band gap is opened by hybridizations between the spin-up
and -down channels. The other subclass of the QAHE has a nonzero spin Chern
number. One representative material is the Mn-doped HgTe \cite{QAHELiu}, where
the $s$-type electrons of Hg and the $p$-type holes of Te experience opposite
$g$-factors when they couple with the $d$ electrons of Mn. The opposite
exchange fields felt by the $s$ and $p$ orbitals enlarge the energy gap in one
spin channel, and close and then reopen the energy gap in the other spin
channel, which induces a nontrivial topology in the latter spin channel due to
the band inversion mechanism~\cite{Bernevig,QAHELiu}, for a large enough
$g$-factor. However, its experimental realization turns out to be difficult
due to the paramagnetic state of Mn spins. Two other materials are
proposed to realize the QAHE with nonzero spin Chern numbers in honeycomb
lattice, a silicene sheet sandwiched by two ferromagnets with magnetization
directions aligned anti-parallelly~\cite{Ezawa2013}, and a perovskite material
LaCrO$_3$ grown along the [111] direction with Cr atoms replaced by Ag or Au
in one atomic layer~\cite{Liang2013,Weng2015}. In both systems, in addition to
the anti-ferromagnetic (AFM) exchange field and spin-orbit coupling (SOC), a
strong electric field is required to break the inversion symmetry in order to
realize the QAHE. For the former one, the weak SOC of silicene limits the
novel QAHE to low temperatures, while for the latter one, growth of the
perovskite material along the [111] direction seems to be difficult.

In the present work, we propose a new material to realize the second subclass
of the QAHE without any extrinsic operation and easy to synthesize. It is
based on LaCrO$_3$ grown along the [001] direction, where we insert one atomic
layer of an inverse-perovskite material Sr$_3$PbO~\cite{SrPbO,Kariyado2012}
such that the Pb atom feels the exchange field established by the Cr atoms in
the parent material. With first-principles calculations, we reveal that there
is a band inversion at the $\Gamma$ point between the $d$ orbital of Cr and
the $p$ orbital of Pb in the spin-up channel induced by the SOC, whereas the
spin-down bands are pushed far away from the Fermi level by the AFM exchange
field. Constructing an effective low-energy Hamiltonian, we explicitly show
that the system is characterized by simultaneous nonzero charge and spin Chern
numbers. Projecting the bands near the Fermi level onto the subspace composed
of the spin-up $d$ and $p$ orbitals by maximally localized Wannier
functions~\cite{MLWF}, we confirm that a spin-polarized and dissipationless
current flows along the edge of a finite sample. Since these two materials are
stable in bulk and match each other with small lattice distortions, the
composite material is expected easy to synthesize.

\section{First-principles calculations}
\begin{figure}[t]
  \centering
  \includegraphics[width=.9\linewidth]{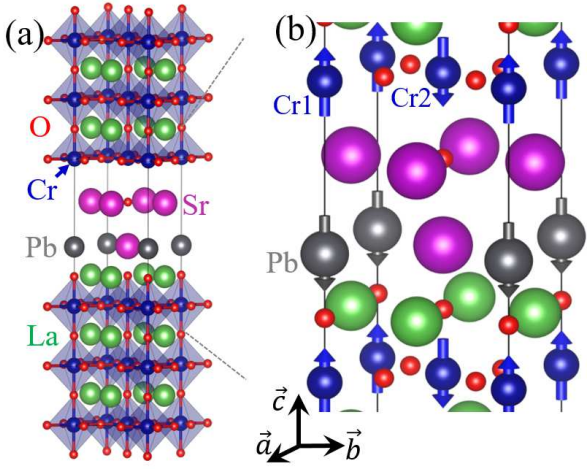}
  \caption{(Color online) (a) Crystal structure of bulk LaCrO$_3$ grown along the [001]
  direction with one atomic layer replaced by Sr$_3$PbO. (b) Enlarged interface
  between LaCrO$_3$ and Sr$_3$PbO with grey and blue arrows representing spin moments on
  Pb and Cr sites, respectively. $\vec{a}$, $\vec{b}$ and $\vec{c}$ are lattice vectors.}
  \label{fig1}
\end{figure}

The parent material LaCrO$_3$ exhibits the perovskite structure with formula
ABO$_3$, where the oxygen atoms form an octahedron surrounding the B atom. It is a
well-known Mott insulator with a large energy gap $\sim$3 eV, carrying the
G-type AFM order, where the spin moment of any Cr aligns opposite to all its
neighbors. On the other hand, the material Sr$_3$PbO shows the
inverse-perovskite structure with formula A$_3$BO, where the A atoms form an
octahedron surrounding the oxygen. It was revealed recently that there is a
topological band gap in bulk Sr$_3$PbO~\cite{Klintenberg2010,HsiehPRB2014}.
We notice that the $\vec{a}$-$\vec{b}$ plane lattice constant is 3.88 \AA~for
LaCrO$_3$, and 5.15 \AA~for Sr$_3$PbO, different from each other by a factor
close to $\sqrt{2}$. Therefore, with a $\pi/4$ rotation around the common
$\vec{c}$ axis, these two materials match each other quite well [see
Figs.~\ref{fig1}(a) and (b)]. At the interface the oxygen of Sr$_3$PbO
completes the CrO$_6$ octahedron of the perovskite structure [see
Fig.~\ref{fig1}(b)], which minimizes the distortion to the two materials when
grown together. As shown in Fig.~\ref{fig1}(b) zoom-in at the interface, there
are two types of Cr atoms in each CrO$_2$ unit cell, where Cr1 sits at the
corners of the square and above the Pb atom in the $\vec{c}$ axis, whereas Cr2
sits at the center of square and above the oxygen.

We have performed first-principles calculations by using density functional
theory (DFT) implemented in the Vienna \textit{Ab-initio} Simulation Package
(VASP) \cite{VASP}, which uses the projected augmented wave (PAW) method
\cite{Blochl,Kresse}. The exchange correlation potential is described by the
generalized gradient approximation (GGA) of Perdew-Burke-Ernzerhof (PBE) type
\cite{PBE}. The cut-off energy of the plane waves is chosen to be $500$ eV.
The Brillouin zone is meshed into a $10\times10\times1$ grid using the
Monkhorst-Pack method. The Hubbard-U term is included for the Cr-$3d$
electrons with $U = 5.0$ eV and $J = 0.5$ eV \cite{Yang1999} by using the
Dudarev method. For the lattice structure, we take $a = b =
5.48$ \AA~$(=\sqrt{2}\times 3.88$ \AA)~and $c_{\rm LaCrO} = 3.88$ \AA.  The
height of the inserted Sr$_3$PbO layer is determined by a relaxation process
to achieve the minimal energy: $c_{\rm SrPbO} = 5.46$ \AA, the distance from
the Pb atom to the Cr just above it (that to the Cr below it is $c_{\rm
LaCrO}$). Afterwards, the positions of atoms are determined by a second
relaxation process with all lattice constants fixed. In both processes, the
criterion on forces between atoms is set to below 0.01 eV/\AA. The results
shown below are for a superlattice structure with five layers of LaCrO$_3$ and
one atomic layer of Sr$_3$PbO. We confirm that the results remain unchanged as
far as the number of LaCrO$_3$ layers is above five and for $U_{\rm eff} = U -
J > 3.5$ eV.

\begin{figure}[t]
  \centering
  \includegraphics[width=8cm]{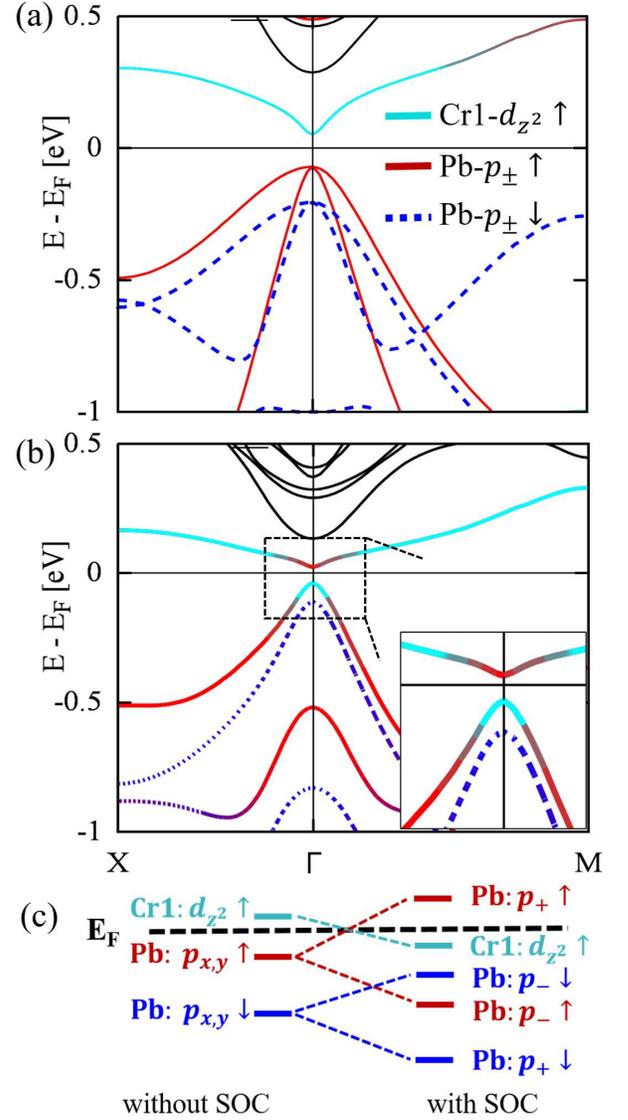}
  \caption{(Color online) Band structure of the supercell shown in Fig.~\ref{fig1} without (a)
  and with (b) SOC. Solid cyan, solid red and dashed blue curves are for
  Cr1-$d_{z^2}^\uparrow$, Pb-$p_\pm^\uparrow$ and Pb-$p_\pm^\downarrow$
  orbitals, respectively. Other colors indicate their hybridizations. (c)
  Schematic band evolution of Pb-$p_{x,y}$ and Cr-$d_{z^2}$ at the $\Gamma$
  point caused by SOC.}
  \label{fig2}
\end{figure}

Without SOC, we find a band gap $0.18$ eV at the $\Gamma$ point.
As shown in Fig.~\ref{fig2}(a), the topmost valence band is
occupied by the spin-up $p_\pm (= p_x\pm ip_y)$ orbitals of Pb, and the lowest
conduction band is contributed by the spin-up $d_{z^2}$ of Cr1. The reason for
this band arrangement is that the Cr1 atom does not live in a closed
octahedron due to the absence of an oxygen in the corner of Sr$_2$O layer as
shown in Fig.~\ref{fig1}(b), which weakens the crystal field splitting and
lowers the energy of the unoccupied Cr1-$d_{z^2}$ band, whereas the Cr2 shares
one oxygen with Sr$_3$PbO, thus is closed by a complete oxygen octahedron,
which keeps its $d_{z^2}$ far away from the Fermi level.  Therefore, only the
spin-up Cr1-$d_{z^2}$ band appears just above the Fermi level, in contrast to
the original Mott insulator. Meanwhile, the Pb acquires a magnetic moment
$0.19\mu_B$ polarized downwards [see Fig.~\ref{fig1}(b)], which matches the
overall AFM order of LaCrO$_3$ and splits the spin-up and spin-down $p$
orbitals of Pb [see Fig.~\ref{fig2}(a)]. In this way, both the topmost valence band
and the bottommost conduction band are occupied by the spin-up channel.
We notice that the total magnetic moment in the
present system is compensated to zero, distinct from the Cr-doped Bi$_2$Se$_3$
\cite{Weng2015,Yu2010}.

The band structure of the material is then calculated with SOC turned on,
which lifts the degeneracy of the $p_+$ and $p_-$ bands in both spin channels.
Remarkably, the strong SOC of the heavy element Pb pushes the $p_+$ orbital
with up spin even above the Fermi energy $E_F$ around the $\Gamma$ point as
displayed in Figs.~\ref{fig2}(b) and (c).  The Cr1-$d_{z^2}$ orbital with up
spin then has to sink across the Fermi level partially in order to maintain
the charge neutrality of the system, which causes a band inversion between the
$p$ and $d$ orbitals around the $\Gamma$ point, as shown in
Fig.~\ref{fig2}(b). An energy gap of $59$ meV is observed according to the
first-principles calculations.

\begin{figure}[t]
  \centering
  \includegraphics[width=.8\linewidth]{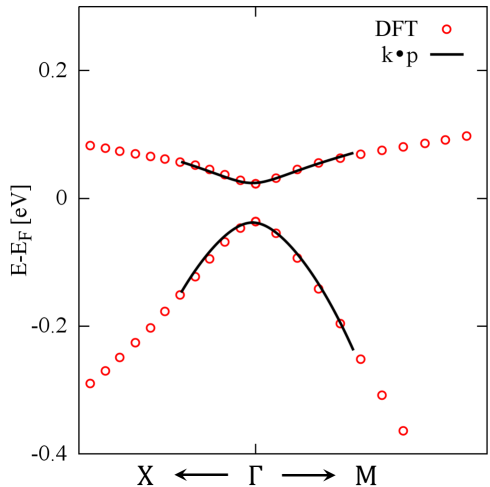}
  \caption{(Color online) Energy dispersion around the $\Gamma$ point fitted
    by the 2$\times$2 $\bk\cdot\bp$ Hamiltonian (\ref{eq:kp}) on the basis $[d_{z^2}^\uparrow,
  p_+^\uparrow]$. The fitted curves collapse with the DFT results within the
  region $|k_x|\leq 0.08\frac{2\pi}{a}$ and $|k_y| \leq 0.08\frac{2\pi}{b}$,
  where $a$ and $b$ are lattice constants given in text.}
  \label{fig3}
\end{figure}

\section{Effective low-energy model}
We now derive an effective low-energy $\bk\cdot\bp$ Hamiltonian to
describe topological properties of the system. Noticing that the topological
band gap is opened by hybridizations between the spin-up $p_+$ orbital of Pb
and the spin-up $d_{z^2}$ orbital of Cr1, it is then reasonable to take these
two orbitals as a basis to construct a 2$\times$2 Hamiltonian. For simplicity,
we denote the two orbitals as $\Gamma_1 = d_{z^2}^\uparrow$ and $\Gamma_2 =
p_+^\uparrow$. The effective $\bk\cdot\bp$ Hamiltonian around the $\Gamma$
point is
\begin{equation}
  H(\bk) = H_0 + H'
\end{equation}
on the basis $[\Gamma_1,\Gamma_2]$, where
\begin{equation}
  H_0 =
  \left(
  \begin{array}{cc}
     \epsilon_1 + \gamma_1\bk^2 & 0 \\
     0 & \epsilon_2 + \gamma_2\bk^2
  \end{array}
  \right)
\end{equation}
and $H' = \bk\cdot\bp =
\left(k_-P_++k_+P_-\right)/2$ is the perturbation term with $k_\pm = k_x\pm
ik_y$ and $P_\pm = P_x\pm iP_y$ ($P_{x/y}$ is the momentum operator in the
$x/y$ direction). Since the crystal is symmetric with respect to the $C_4$ rotation around
the $\vec{c}$ axis, $H'$ must be invariant under the $C_{4} =
e^{-i\frac{2\pi}{4}J_z}$ transformation, where $J_z$ the is the $z$-component of the
total angular momentum. The symmetry constraint allows us to
determine nonzero entries of $H'$ \cite{Yu2015}. It is easy to check that
$C_{4}\Gamma_1 = e^{-i\frac{\pi}{4}}\Gamma_1$ and $C_{4}\Gamma_2 =
e^{-i\frac{3\pi}{4}}\Gamma_2$ because $J_z = 1/2$ and $3/2$ for $\Gamma_1$ and
$\Gamma_2$ respectively. Since
\begin{eqnarray}
\left<\Gamma_1|P_+|\Gamma_2\right>&=&\langle\Gamma_1|C_4^\dagger
  C_4 P_+ C_4^\dagger C_4|\Gamma_2\rangle \nonumber\\
  &=&\langle\Gamma_1|e^{i\frac{\pi}{4}} e^{-i\frac{\pi}{2}} P_+
  e^{-i\frac{3\pi}{4}}|\Gamma_2\rangle \nonumber \\
  &=&-\left<\Gamma_1|P_+|\Gamma_2\right>,
\end{eqnarray}
$\left<\Gamma_1|k_-P_+|\Gamma_2\right>$ must vanish. Performing similar
calculations for all other terms, we arrive at the Hamiltonian
respecting the crystal symmetry
\begin{equation}
  H(\bk) = \left(\epsilon_0 + \gamma_0 k^2\right)I_{2\times2} +
  \left(
  \begin{array}{cc}
    \epsilon+\gamma \bk^2 & \alpha k_+ \\
    \alpha^* k_- & -\epsilon-\gamma \bk^2 \\
  \end{array}
  \right)
  \label{eq:kp}
\end{equation}
up to the lowest orders of $\bk$, with
$\epsilon_0=(\epsilon_1+\epsilon_2)/2$, $\epsilon=(\epsilon_1-\epsilon_2)/2$,
$\gamma_0=(\gamma_1+\gamma_2)/2$ and $\gamma=(\gamma_1-\gamma_2)/2$. By
fitting the energy dispersion of $H(\bk)$ in Eq.~(\ref{eq:kp}) against
the first-principles results given in Fig.~\ref{fig2}(b), we obtain the parameters
as follows: $\epsilon_0=-0.007$ eV, $\gamma_0 = -7.8$ eV$\cdot$\AA$^2$,
$\epsilon=-0.031$ eV, $\gamma=9.0$ eV$\cdot$\AA$^2$ and $\alpha = 1.45$ eV$\cdot$\AA~(see
Fig.~\ref{fig3}). Since $\epsilon$ and $\gamma$ take opposite signs,
the electronic wavefunction of the spin-up channel becomes topologically nontrivial
due to the band inversion mechanism with Chern number $C_\uparrow =1$. Since the
spin-down electronic bands are kept far away from the Fermi level [see
Figs.~\ref{fig2}(b) and (c)], one clearly
has $C_\downarrow = 0$. It is therefore confirmed that the system is
characterized by simultaneous charge and spin Chern numbers: $C_c = C_\uparrow
+ C_\downarrow = 1$ and $C_s = C_\uparrow - C_\downarrow = 1$.

\begin{figure}[t]
  \centering
  \includegraphics[width=.8\linewidth]{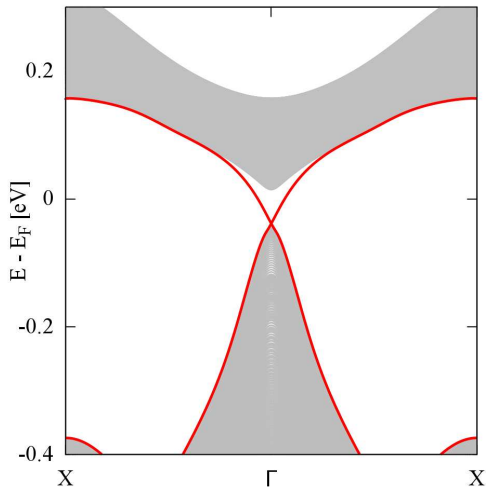}
  \caption{(Color online) Band structure for a slab of the system shown in
  Fig.~\ref{fig1} based on Wannierized wavefunctions downfolded from the
  results of the first-principles calculations in Fig.~\ref{fig2}(b).  Red
  curves are for topological edge states in the spin-up channel, and grey ones
  are for bulk states.}
  \label{fig4}
\end{figure}

\section{Topological edge states}
The nontrivial topology gives rise to gapless edge states in a finite sample.
To illustrate this feature, we calculate the dispersion relation for a slab of
the topological material with $100a$ along the $\vec{a}$ axis and infinite
along the $\vec{b}$ axis (see Fig.~\ref{fig1}). Since the bulk bands close to the
Fermi level are mainly contributed by the Pb-$p_x$, Pb-$p_y$ and Cr1-$d_{z^2}$
orbitals, it is reasonable to downfold the wavefunctions obtained by the
first-principles calculations in Fig.~\ref{fig2}(b) onto these three orbitals.
Employing the maximally-localized Wannier functions \cite{MLWF}, we obtain the
hopping integrals within the six-dimensional subspace including the spin
degree of freedom. It is then straightforward to calculate the band structure
of the slab system.  As shown in Fig.~\ref{fig4}, a gapless edge state with up
spin appears inside the bulk gap, manifesting the nontrivial topology of the
present system.

\section{Discussion}
Liu \cite{Liu2013} proposed a 3D spinless model for a layered square lattice
with A-type AFM (intra-plane ferromagnetic and inter-plane AFM orderings). At
each lattice site, there are three orbitals: $s$, $p_x$ and $p_y$. Each layer
can be driven into a QAHE in a same way as that for the Mn-doped HgTe
\cite{QAHELiu}. Since every two adjacent layers have opposite magnetic
moments, their chiral edge states propagate counter to each other.
Therefore, the system can be viewed as a stack of quantum spin Hall
insulators [see also \cite{MongPRB2010}], where the combination of the
time-reversal and the primitive-lattice translational symmetries is
preserved. In contrast, all symmetries are broken in our system, giving rise
to a Chern insulator.

\section{Conclusion}
We propose a novel topological material composed of the LaCrO$_3$ of perovskite
structure grown along the [001] direction with one atomic layer replaced by
an inverse-perovskite material Sr$_3$PbO. Based on first-principles
calculations and an effective low-energy Hamiltonian, we demonstrate that the
topological state is characterized by simultaneous nonzero charge and spin
Chern numbers, which can support a spin-polarized and dissipationless edge current
in a finite sample. Supported by the anti-ferromagnetic exchange field and
spin-orbit coupling inherent in the compounds, no extrinsic operation is
required for achieving the novel topological state. Importantly, these two materials
are stable in bulk and match each other with only small lattice distortions, which
makes the composite material easy to synthesize.

\section*{Acknowledgments}
This work was supported by the WPI Initiative on Materials Nanoarchitectonics,
Ministry of Education, Culture, Sports, Science and Technology of Japan. QFL
acknowledges support from the National Natural Science Foundation of China
(No.11574215) and the Scientific Research Foundation for the Returned Overseas
Chinese Scholars, State Education Ministry.


\end{document}